\documentclass[twocolumn,showpacs,preprintnumbers,amsmath,amssymb]{revtex4}

\usepackage{graphicx}
\usepackage{dcolumn}
\usepackage{bm}
\usepackage[usenames]{color}

\begin{document}

\preprint{APS/123-QED}

\title{Density imbalance effect on the Coulomb drag upturn in an undoped electron-hole bilayer}

\author{Christian P. Morath}\email{cpmorat@sandia.gov}
\altaffiliation[Also at ]{Physics and Astronomy Department, University of New Mexico.}

\author{John A. Seamons}%

\author{John L. Reno}%

\author{Michael P. Lilly}

\affiliation{Sandia National Laboratories, Albuquerque, NM 87185
}%

\date{\today}

\begin{abstract}
A low-temperature upturn of the Coulomb drag resistivity measured in an undoped electron-hole bilayer (uEHBL) device, possibly manifesting from exciton formation or condensation, was recently observed.  The effects of density imbalance on this upturn are examined.  Measurements of drag as a function of temperature in a uEHBL with a 20 nm wide Al$_{.90}$Ga$_{.10}$As barrier layer at various density imbalances are presented.  The results show drag increasing as the density of either two dimensional system was reduced, both within and above the upturn temperature regime and with a stronger density dependence than weak-coupling theory predicts.  A comparison of the data with numerical calculations of drag in the presence of electron-hole pairing fluctuations, which qualitatively reproduce the drag upturn behavior, is also presented.  The calculations predict a peak in drag at matched densities, which is not reflected by the measurements.
\end{abstract}




\maketitle


An exciton is a composite boson that forms in bulk semiconductors due to an attractive Coulomb interaction between its fermionic, constituent electron and hole.  As such, excitons are expected under certain circumstances to condense at low temperature, where the lowest energy state becomes occupied by a macroscopic number of particles.  While the bulk exciton condensate was later determined to be an insulator due to interband transitions which fix the phase of the order parameter\cite{Guseinov1972}, the use of spatially-separated electron-hole pairs or "indirect" excitons was predicted to mitigate this issue sufficient for a phase transition to occur.\cite{Lozovik1975, Lozovik1976}

Indirect excitons may be generated optically\cite{Butov1994} or via field-effect\cite{Seamons2007, Keogh2005} in double quantum wells.  The distinct advantages of field-effect devices, such as the uEHBL used in this study, are that the densities in each well can be adjusted and then maintained at constant values using gate voltages and the layers have separate electrical contacts to each.  Together these allow for the interlayer Coulomb interaction between the electrons and holes to be probed directly using Coulomb drag measurements.  Conceived of by Progrebinsky\cite{Pogrebinskii1977} and Price\cite{Price1983} and first demonstrated between two dimensional electron gases (2DEGs) by Gramila \textit{et al.}\cite{Gramila1991}, in the Coulomb drag technique a current is driven in one layer of a bilayer device causing a longitudinal voltage to arise in the adjacent layer via interlayer scattering.  The measured quantity is the drag resistivity $\rho_{D} = V_{drag}/I_{drive}(L/W)$, where $I_{drive}$ is the current in the drive layer, $V_{drag}$ is the induced voltage in the drag layer and $L/W$ is the number of squares. In the "weakly-coupled" limit, low temperature $T$ and large interlayer separation $d$, the $\rho_{D}$ is expected to have a $T^{2}$-dependence, due to phase space restrictions on the scattering set by the thermal broadening, and thereby decrease to zero as $T \rightarrow 0$.\cite{Gramila1991,Jauho1993}  Deviation of $\rho_{D}$ from this behavior, possibly due to enhanced interlayer coupling, would thus suggest a departure from Fermi-liquid physics.



Seizing upon this possibility, Vignale and MacDonald predicted that $\rho_{D}$ in an electron-hole bilayer system with a superfluid condensate would jump discontinuously at the condensation temperature $T_{C}$ and diverge as $T \rightarrow 0$.\cite{Vignale1996}  In their theory, the current was partitioned into a superfluid portion carried by the condensate and a normal portion carried by the quasiparticles.   Further theoretical work by Joglekar \textit{et al.}\cite{Joglekar2005}, which treated the system as a dipolar condensate, confirmed the expected divergence in $\rho_{D}$ as a consequence of the reduction in the quasiparticle density and the consequently larger electric field required to drive the normal component of the current.  Hu also predicted an enhancement of $\rho_{D}$ above $T_{C}$ due to electron-hole pairing fluctuations.\cite{Hu2000}  This mechanism, which is analogous to short-lived Cooper pairs in superconductors, is discussed further below.  Thus, any evidence of electron-hole pairing in a bilayer device is expected to manifest in $\rho_{D}$ measurements as a function of $T$.

Condensate formation in bilayers was also predicted to manifest as a supercurrent\cite{Lozovik1975}; however, new theory predicts additional restrictions on the experimental setup for observing this supercurrent.\cite{Su2008}  For any pairing to occur, however, a requirement for devices with $d \leq n^{-1/2}$, where $n^{-1/2}$ is the typical interparticle distance of the two dimensional system (2DS) with density $n$, is expected.\cite{Balatsky2004}  Practically speaking, such devices are difficult to fabricate and this, in turn, has made finding an electron-hole condensate in a bilayer an elusive goal.

The first measurements of $\rho_{D}$ in an electron-hole bilayer were accomplished almost two decades ago by Sivan \textit{et al.}\cite{Sivan1992} and exhibited behavior characteristic of weakly-coupled 2DSs dominated by Coulomb scattering. Recently, however, electron-hole bilayer devices with thinner barrier layers $(\leq20$ nm$)$ and lower densities ($<10^{11}$ cm$^{-2}$) were produced\cite{Seamons2007b, DasGupta2008, Croxall2008, Seamons2008} and deviations from the weak-coupling $T^2$ drag behavior began emerging.  Early indications came from Seamons\cite{Seamons2007b}, where a distinct upturn of $\rho_{D}$ measured in the hole layer was found at $T\sim .5$ K in two 20 nm barrier width samples.  No upturn in $\rho_{D}$ measured in the electron layer was found, however, possibly because of self-heating from driving current through the highly resistive two dimensional hole gas (2DHG).  Self-heating also precluded measuring $\rho_{D}$ of the electron layer in this work.



Similar results were concurrently found by the Cambridge group\cite{DasGupta2008,Croxall2008}, who also highlighted how the difference in $\rho_{D}$ from interchanging the drag and drive layers directly contradicts the Onsager reciprocity theorem.  It was subsequently shown that the $\rho_{D}$ upturn was followed by a downturn and saturation at a small negative value.\cite{Croxall2008}  Finally, a direct relationship between $T_{U}$, the temperature at which the minimum in $\rho_{D}$ occurs, and matched electron and hole densities $n=p$ was revealed.\cite{Seamons2007b,Seamons2008}  While the details of the $\rho_{D}$ upturn phenomena remain speculative, exciton formation or condensation is often conjectured to be its source.  Beginning to examine this conjecture using the simple means of density imbalance is the primary goal of this Communication.

Here the effects of density imbalance on the low temperature upturn of $\rho_{D}$ in a uEHBL are reported.  The $\rho_{D}$ was measured as a function of $T$ for various unmatched densities $n\neq p$ in both 2DSs of the uEHBL.  The data showed that $\rho_{D}$ increased as the density of either 2DS was reduced, with a stronger density dependence than weak-coupling theory predicts.  Numerical calculations of electron-hole pairing fluctuation theory were also done for similar density imbalances.\cite{Hu2000}  While the calculations qualitatively reproduced the upturn observed in the measurements, they also predicted a peak in $\rho_{D}$ centered at $n=p$, which was not observed.

\begin{figure}
\begin{center}
\includegraphics[width = 9.5cm]{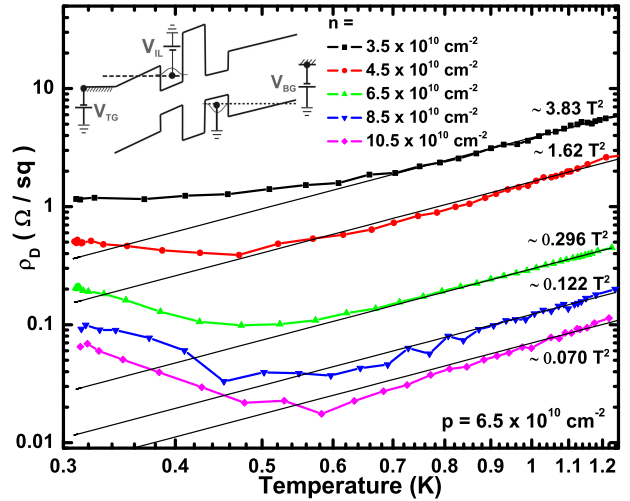}
\end{center}
\caption{\label{fig1}(Color online) Upturn in $\rho_{D}$ measured as function of $T$ for $n$ ranging from $3.5$ to $10.5 \times 10^{10}$ cm$^{-2}$ at $p=6.5 \times 10^{10}$ cm$^{-2}$.  Thin black lines are $T^{2}$ best fits.  Inset shows a schematic of the bandstructure during operation.}
\end{figure}

The details of fabricating and operating uEHBLs were previously discussed.\cite{Seamons2007, Morath2008, Seamons2008}  Based on these results, Hwang and Das Sarma determined the 2DS's mobility in uEHBLs was background charged impurity scattering limited and the enhancement of $\rho_{D}$ well above $T_{C}$ was due to exchange effects.\cite{Hwang2008}  A schematic depicting the bandstructure of a uEHBL during operation, including the top-gate voltage $V_{TG}$, interlayer voltage $V_{IL}$, and back-gate voltage $V_{BG}$, is shown in the inset of Fig. \ref{fig1}.  The $n$ and $p$ are predominantly determined by $|V_{TG}- V_{IL}|$ and $V_{BG}$, respectively.  The sample (EA1287 6.3) used in this study had a 20 nm wide Al$_{.90}$Ga$_{.10}$As barrier separating 18 nm GaAs quantum wells.  The $n$ and $p$ were measured simultaneously at $T = 0.3$ K prior to each $\rho_{D}$ temperature sweep using low-field Hall measurements by standard ac lock-in technique with 10 nA drive currents in each 2DS.  The $\rho_{D}$ measurements were also performed with standard ac lock-in technique using a 50 nA drive current in the 2DEG at 3.5 Hz.  Since the 2DEG is held at $V_{IL}=-1.465$ V, the current is coupled in via an isolation transformer. At this $V_{IL}$ the inter-layer leakage current was $\sim 1$ nA and, as discussed in \cite{Seamons2008}, had no discernable effect on the upturn.

Measurements of the upturn in $\rho_{D}$ at $p = 6.5 \times 10^{10}$ cm$^{-2}$ for various drive layer densities $n$ are given in Fig. \ref{fig1}. The black lines are best-fits $A \cdot T^{2}$, where $A$ is the single fitting constant and $T^{2}$ is the characteristic temperature dependence that results from phase-space requirements in weak-coupling Fermi-liquid theory.\cite{Jauho1993}  Similar results were found for $\rho_{D}$  measurements at $n = 8.5 \times 10^{10}$ cm$^{-2}$ for various drag layer densities $p$.  Summarizing the behavior, the fit lines provide a clear indication that for $T>T_{U}$ the $\rho_{D}$ followed the expected $T^{2}$-dependence for Coulomb scattering of a weakly-coupled 2DEG and 2DHG.  The data also adhered to the following weak-coupling predictions: (1) at $p=n$, the $\rho_{D}$ increased as matched density was reduced; and, (2) for $p \ne n$, the $\rho_{D}$ increased if either density was reduced.


In the upturn regime, $T\leq T_{U}$, the following behaviors are visible: (1) at $p=n$, the $T_{U}$ increased as total density $n + p$ was increased, similar to what was previously reported \cite{Seamons2008}; and, (2) for $p \ne n$, the $T_{U}$ also increased as either $p$ or $n$ was increased.  Fig.\ref{fig1} also indicates the upturn is most strongly dependent on $T$ at $n = 10.5 \times 10^{10}$ cm$^{-2}$ and becomes comparitively weaker as $n$ decreases, eventually showing a saturation behavior at $n = 3.5 \times 10^{10}$ cm$^{-2}$.


\begin{figure}
\begin{center}
\includegraphics[width = 9.5cm]{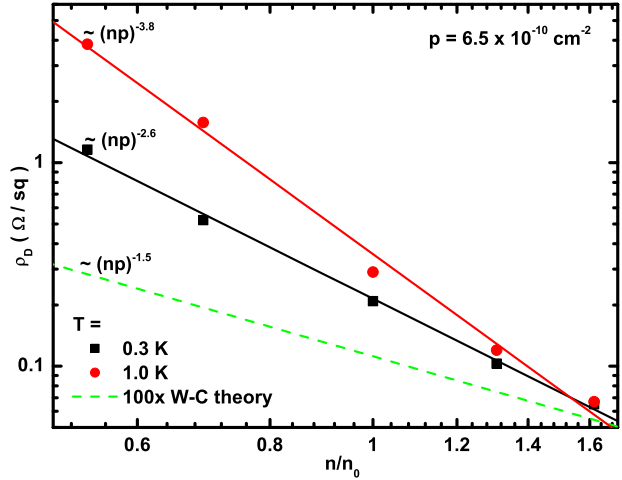}
\end{center}
\caption{\label{fig2}(Color online) $\rho_{D}$ as a function of $(n/n_{0})$, where $n_{0}= p = 6.5 \times 10^{10}$ cm$^{-2}$, at $T=0.3$ and $1.0$ K.  Dotted line is $100\times$ weak-coupling analytic theory at $T=0.3$ K.}
\end{figure}


In Fig. \ref{fig2}, the same $\rho_{D}$ data from Fig. \ref{fig1} at $T=0.3$ and $1.0$ K is plotted as a function of $(n/n_{0})$, where $n_{0}=p= 6.5 \times 10^{10}$ cm$^{-2}$.  The dotted line in Fig. \ref{fig2} is calculated using the analytic expression for $\rho_{D}$, which applies in the limit of large layer spacing $d$ and for low $T$, given by $\rho_{D} = \alpha T^{2}/(np)^{3/2}d^{4}$, where $\alpha = \hbar \xi(3)(4 \pi \kappa \varepsilon_{0} k_{B})^{2}/128 \pi e^{6}$.\cite{Jauho1993}  Here $\hbar$ is Planck's constant,   $\xi(3) \sim 1.202$ is the Riemann zeta function, $\kappa$ is the dielectric constant of GaAs, $\varepsilon_{0}$ is permittivity of free space and $k_{B}$ is Boltzmann's constant.  The weak-coupling theory is known to dramatically underestimate the measurements\cite{Seamons2007} and, to aid in the comparison, the dotted line is $100\times$ the theoretical results at $T = .3$ K.

Summarizing, the main result from Fig. \ref{fig2} is the monotonic decrease of the measured $\rho_{D}$ as $np$ was increased, both above and within the upturn regime.  This decrease in $\rho_{D}$ was consistent through the $n=p$ case and for both varying $n$ and $p$ measurements (latter is not shown).  As discussed further below, this monotonic decrease with $np$ does not follow the predicted behavior for $\rho_{D}$ in the upturn regime, where a peak at $n=p$ was predicted.\cite{Hu2000}

Additionally, the log-log plot in Fig. \ref{fig2} also allows for a direct comparison of the $\rho_{D}(np)$-dependence in each regime.  Weak-coupling theory predicts $(np)^{-3/2}$, as shown above.  The measurements, however, roughly follow $(np)^{-2.9}$ and $(np)^{-3.7}$ at $T=0.3$ and $1.0$ K, respectively.  These exponents are both larger than $\sim 1.8$, which was predicted\cite{DasSarma2008} for this uEHBL based on the theory in \cite{Hwang2008}.  Larger exponents were also previously observed in both 2DHG-2DHG and 2DEG-2DEG drag.\cite{Pillarisetty2002,Kellogg2002b}

\begin{figure}
\begin{center}
\includegraphics[width = 9.5cm]{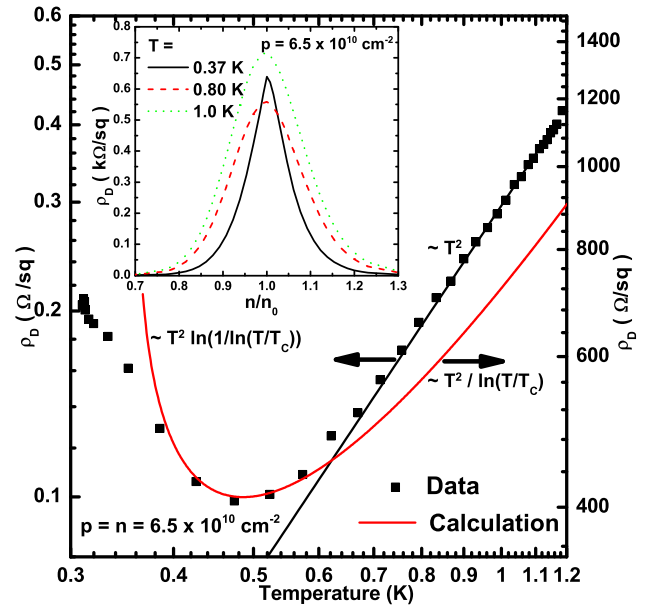}
\end{center}
\caption{\label{fig3}(Color online) Results of pairing-fluctuation calculations of $\rho_{D}$ as a function of $T$ plotted alongside measured data for $p=n=6.5 \times 10^{10}$ cm$^{-2}$. Solid black line is a $T^{2}$ fit to the measured data. Inset: Calculations of $\rho_{D}$ plotted as a function of $(n/n_{0})$ at $T=0.37$, $0.80$ and $1.0$ K where $n_{0}= p = 6.5 \times 10^{10}$ cm$^{-2}$ and $n$ was varied from $4.5$ to $9.0\times 10^{10}$ cm$^{-2}$.}
\end{figure}


To begin examining the experimental results above, a comparison to numerical calculations of Hu's drag equation is made in the following.\cite{Hu2000} The reason most often quoted for the upturn in $\rho_{D}$ is electrons and holes entering a paired state\cite{Seamons2007b, Croxall2008, Seamons2008}, as anticipated by Vignale \textit{et al.} \cite{Vignale1996}, Hu \cite{Hu2000} and Balatsky \textit{et al.}\cite{Balatsky2004}  The drag equation devised by Hu, however, offers the simplest means to begin appraising the density imbalance effect on the drag upturn observed in the experimental data.  Hu's pairing fluctuation analysis indicates $\rho_{D}$ will be significantly enhanced above the mean field transition temperature $T_{C}$, similar to the effect of ephemeral Cooper pairs on the conductivity above $T_{C}$ in superconductors.  The calculation neglects to account for impurity potentials and bandstructure effects.  It uses a simple local interlayer interaction $V(q)=V_{0}$, which, unlike the more realistic Coulomb interaction\cite{Jauho1993, Yurtsever2003}, fails to cut off the large momentum transfer contributions and thereby significantly overestimates the drag.  Despite this well-understood shortcoming, the pairing fluctuation analysis provides the only qualitative comparison for the upturn in $\rho_{D}$ with the density imbalance data.

An example of a $\rho_{D}$ calculation is shown in Fig. \ref{fig3}, alongside measured results at $p=n=6.5 \times 10^{10}$ cm$^{-2}$ from Fig. \ref{fig1}.  For this curve $T_{C}=.36$ K was chosen by hand so that $T_{U}$ of the calculated curve would best match the $n=p$ data.  The measured data and the calculated curve show qualitatively similar \textit{non-monotonic} dependencies on temperature; both traces show $\rho_{D}$ decreasing with $T$ and then abruptly upturning at $T_{U}$.  However, the calculated curve predicts a drag magnitude $3$ orders larger than the measured data.  It also has different temperature dependencies than the data for both $T \leq T_{U}$ and $T>T_{U}$.   In the former, the measured data is finite, while the calculations follow a $T^{2}ln(1/ln(T/T_{C}))$-dependence, which diverges. For $T>T_{U}$, the calculations follow $T^{2}/ln(T/T_{C})$-dependence, which differs from the $T^{2}$-dependence of the data, indicated by the thin, black line in Fig. \ref{fig3}.


The $T_{C}$ for the calculated $\rho_{D}$ curves at $n\neq p$ were determined according to the following procedure.  For $n<p$ the $T_{C}= .36(n/p)$ K.  For curves at $p<n$ the $T_{C}=.36$ K was used.  This procedure assumes the density of excitons $n_{ex}$ is some fraction of the lesser of $n$ and $p$ and that the transition temperature is proportional to the density, in accordance with the discussion in \cite{Butov2004}.

Calculated results at $T=0.37$, $0.8$, and $1.0$ K are plotted in in the inset of Fig. \ref{fig3} as a function of $(n/n_{0})$, where $n_{0}=p=6.5 \times 10^{10}$ cm$^{-2}$ and $p$ was held constant while $n$ was varied from $4.5$ to $9.0 \times 10^{10}$ cm$^{-2}$.  These results predict $\rho_{D}$ is sharply peaked at $n=p$ for temperatures within and above the upturn regime $(T>.5$ K$)$, in stark contrast to the measured results in Fig. \ref{fig2}, where $\rho_{D}$ increased monotonically with decreasing density.




Thus, while it appears from Fig. \ref{fig3} that measured data has a qualitatively similar nonmonotonic temperature dependence to predictions based on pairing fluctuations, the results in Fig. \ref{fig2} and the inset of Fig. \ref{fig3} indicate a sharp difference in their dependence on density imbalance.  On the surface, this suggests the $\rho_{D}$ upturn phenomena observed in the measured results is not a manifestation of electron-hole pairing fluctuations above $T_{C}$.



In conclusion, the effects of density imbalance on the low temperature upturn in $\rho_{D}$ of a uEHBL were investigated using Coulomb drag measurements.  Reducing either 2DS density was found to increase $\rho_{D}$ for $T\leq T_{U}$ and $T>T_{U}$.  In each regime $\rho_{D}$ also had stronger $np$-dependence than what's predicted by weak-coupling theory.  While calculations of $\rho_{D}$ in the presence of electron-hole pairing fluctuations were qualitatively able to reproduce the measured upturn behavior, the predicted a peak in $\rho_{D}$ at $n=p$ that was absent from the measured data.



\begin{acknowledgments}
\label{Section5}
It a pleasure to acknowledge technical assistance from Denise Tibbetts and discussions with Sankar Das Sarma, Euyheon Hwang, Ben Hu, Sasha Balatsky and Dan Huang. This work has been supported by the Division of Materials Sciences and Engineering, Office of Basic Energy Sciences, U.S. Department of Energy. Sandia is a multiprogram laboratory operated by Sandia Corporation, a Lockheed Martin Company, for the United States Department of Energy under Contract No. DE-AC04-94AL85000.
\end{acknowledgments}


\bibliography{EHBL_Bib}

\end{document}